\documentclass{article}  

\let\ssection=\section
\renewcommand{\section}{\setcounter{equation}{0}\ssection}
\usepackage{graphicx}
\usepackage{psfrag}
\usepackage{amsmath,amssymb,amsfonts}

\title{Is general relativity `essentially understood' ?
\footnote{Extended version of a talk which was to be delivered at the DPG
Fr\"uhjahrstagung in Berlin, 5 March 2005.} }
\author{Helmut Friedrich\\ 
Max-Planck-Institut f\"ur Gravitationsphysik\\
Am M\"uhlenberg 1\\
14476 Golm\\
Germany}

\begin{document}
\maketitle

\begin{abstract}

\noindent
The content of Einstein's theory of gravitation is encoded in the properties
of the solutions to his field equations. There has been obtained a wealth of
information about these solutions in the ninety years the 
theory has been around. It led to the prediction and the observation of
physical phenomena which confirm the important role of general
relativity in physics. The understanding of the domain of highly dynamical,
strong field configurations is, however, still quite limited. The
gravitational wave experiments are likely to provide soon observational data
on phenomena which are not accessible by other means.   Further theoretical
progress will require, however, new methods for the analysis and the
numerical calculation of the solutions to Einstein's field equations on
large scales and under general assumptions.   We discuss some of the
problems involved,  describe the status of the field and recent results, and
point out some  open problems. 

\end{abstract}




\section{Introduction}

The fascinating reports on the exciting new theories, which propose to
unite our present and all future ideas about space-time, gravitation, and
quantum physics into one coherent scheme from which general
relativity will be derived in the end as a particular limit, may
suggest that Einstein's classical general relativity is essentially
understood. This impression is easily corroborated by the amazing successes
of general relativity. Einstein concludes his synopsis from 1916 of the
theory of general relativity (\cite{einstein:1916a}) with three
predictions: the red shift, the bending of light rays, and the
precession of the perihelion in planetary motion. In view of the
present observational facts the revolutionary character these
implications may have had at the time falls into  oblivion. The red
shift can be measured to day in terrestial experiments
(\cite{pound:rebka}), the bending of light rays is used under the
heading `gravitational lensing' as an effective astrophysical tool
(\cite{brainerd:kochanek}), and various double
star systems provide laboratories for relativistic gravity 
which let the minute advance of the periastron of Mercury 
of 42.98 arcsec/century fade into insignificance: in the case of the
binary pulsar PSR 1913 + 16 the observed advance of periastron is
4.226595(5) deg/yr (\cite{weisberg:taylor}) and for the recently
discovered double pulsar J0737-3039 it even amounts to 16.9 deg/yr
(\cite{kramer:etal}).

In the articles \cite{einstein:1916b}, \cite{einstein:1918}
Einstein discusses `gravitational radiation' and
derives the famous quadrupole formula. Doubts have been raised
subsequently whether the notion of gravitational radiation
referred to a real physical phenomenon (cf. \cite{pais}), but
again the prediction has been confirmed convincingly. Using
Einstein's quadrupole formula to  calculate the rate of period
decrease of the system PSR 1913 + 16 due to its emission of
gravitational radiation, one obtains, after taking into account
certain small corrections, a curve which shows an uncanny
agreement (to within about 0.2 percent) with the data gathered
over  the last 30 years (\cite{weisberg:taylor}). 

The global studies of general relativity starting in the second
half of the last century, led, together with unexpected observations,
to concepts which went far beyond what had been envisioned by
Einstein. Among those the notion of a black hole, a pure space-time
structure which has no place in special relativity, is
certainly the most remarkable one. Though the derivation of detailed 
observational information still poses difficulties, the present
situation suggests that black holes have to be accepted
as part of our reality (\cite{chrusciel:2002}, \cite{rees:2003}).
More  could be said in support of general relativity but I shall leave it at
that.

The overwhelming success of general relativity alluded to above 
may suggest a clear and simple answer to the question posed in
the title of the present article. Also, we understand, of course, 
what it means that the geometry of the world is modelled  
by a Lorentz metric $g_{\mu \nu}$ on a $4$-dimensional
manifold $M$ and that this structure is governed
by Einstein's field equations
\begin{equation}
\label{einst}
R_{\mu \nu} = \kappa\,(T_{\mu \nu} - \frac{1}{2}\,T\,g_{\mu \nu}) +
\lambda\,g_{\mu \nu},
\end{equation}
with cosmological constant $\lambda$, together with equations for
matter fields which define the energy-momentum tensor $T_{\mu \nu}$ and its
trace $T$. In fact,  the predictions referred to above have been derived
from these requirements. There is, however, still a large and potentially
most important part of the theory we do not have access to, neither
mathematically, nor theoretically, nor observationally. 

On the observational side this situation may change soon. With
the gravitational wave experiments presently becoming operational,
we may well enter one of the most important eras of experimental
relativity. The mathematical or theoretical side seems to be lacking
behind, however. In spite of a few general and a rich collection of
specific results our qualitative and
quantitative understanding of {\it highly dynamical processes, strong
field situations, and Einstein evolution over long time scales under
general assumptions} is still quite limited. 

This may sound more 
pessimistic than it should. The field has seen a
substantial progress in recent years and for young researchers there
is a unique opportunity to contribute to the theoretical investigation
of an unexplored domain of fundamental physics. Confronted with the  
gravitational wave measurements these studies may even result in the
discovery of new phenomena which might give missing clues in which
directions to search in the development of `new theories'.

Because the content of the theory is essentially defined by
Einstein's equations it is not surprising that the main problem in the
field is the analysis of these equations and their solutions. 
The following is meant to give an outline of the present situation
and some insight into the questions to be dealt with. It
is definitely not to be considered a survey of a field which has been
growing since 1915.  When I try to explain the situation I shall avoid
technicalities to the extend to which this is possible without getting
too vague in a field which abounds with technical questions by its
very nature. References will be given mainly to illustrate a point or
to direct the reader to more precise statements about concepts and
results. The given references to specialized survey articles should
allow the reader to get a more complete picture.

\section{Gravitational radiation, singularities, and black holes}
\label{radsingbh}

Einstein's analysis of gravitational radiation was based on 
the linearized field equations and the quadrupole formula. As indicated
above, this gives essentially correct results in the `weak' field of the
binary pulsar and it is in fact still the basis for many calculations
of gravitational wave emission in gravitational collapse
scenarios. It can, however, hardly be expected to provide reliable
answers in situations involving strong and highly dynamical fields,
it is useless for calculating radiation generated by the coalescence of
black holes. 

The reconsideration of the idea of gravitational
radiation in  the non-linear theory led in the 1960's to a concept of
gravitational radiation, which does not require mathematical
approximations but relies on the idealization of {\it asymptotic
flatness} in null directions  (\cite{bondi:et.al}). The latter is based
on a picture of the overall behaviour of the
gravitational field of an isolated self-gravitating system which assumes
the field to become weaker and weaker in a characteristic way along null
geodesics running out to null infinity. One might expect that `close to null
infinity linearized gravity takes over', but the situation turned out to be
more subtle than that.  Penrose has given a geometric and particular
elegant characterization of asymptotic flatness in terms of the
extendibility of the conformal structure through null infinity with a
certain degree of smoothness (\cite{penrose:1963}). This suggestion
associates with the far field of a selfgravitating isolated
system a  natural concept of {\it radiation field} but it poses at the
same time difficult questions about the long time behaviour of
gravitational fields and the nature of the field equations. It will be
referred to in the following as  {\it Penrose proposal} (cf.
\cite{frauendiener}, \cite{friedrich:tueb} for further discussion and
references).

As it turned out, this was only a first realization of the
importance of global considerations in the analysis of the
non-linear field equations and the introduction of various general
relativistic concepts. The dominating topics of the following years
were concerned with physical situations such as {\it
gravitational collapse} and {\it cosmological singularities}, were
fields could be expected to show even unlimited growth over finite 
times. Some of the main results of that period of research were (i) the
singularity theorems of Penrose \cite{penrose:1965} and Hawking and
Penrose
\cite{hawking:penrose}, which showed that the occurrence of
space-time singularities is a stable feature of gravitational
fields, (ii) the analysis by Belinskii, Khalatnikov, and Lifshitz
\cite{belinskii:khalatnikov:lifshitz:1970},
\cite{belinskii:khalatnikov:lifshitz:1982}, which led to the
{\it BKL conjecture} about the general (oscillatory) behaviour of the
field near cosmological singularities, and (iii) the development of the
theory of black holes (\cite{frolov:novikov}, \cite{hawking:ellis}).  

It may be noted that `singularity' is defined in the singularity
theorems as the existence of a causal
geodesic which is non-extendible and non-complete. It is left open,
for which reason the curve should be non-extendible. In the case of
specific explixit solutions more could be said concerning the nature
of their singularities 
(cf. \cite{hawking:ellis} for the discussion of examples)
but in general the methods given at the time
were not sufficient to supply more  information.  

Among the questions raised by gravitational collapse theory the
most important and still unsolved one is whether the
evolution of gravitational fields admits a {\it cosmic censorship
principle} (\cite{penrose:1969}) which excludes in {\it generic}
circumstances the existence of {\it naked singularities}, i.e.
singularities which could be seen by (possibly distant) observes. 
If such singularities existed stably under small perturbations this
would reduce the predictive power
of the theory. Giving a precise meaning to this principle is part of
the problem. 
There exist different formulations in the literature such as the {\it
weak cosmic censorship principle}, which asserts that singularities
are hidden within black holes (\cite{wald:cosc}), and
the {\it strong cosmic censorship principle}, which asserts that
maximally extended space-times arising from {\it generic} non-singular
initial data are globally hyperbolic (a notion explained below)
(\cite{moncrief:eardley}, \cite{penrose:1979}). 

There have been given fascinating accounts of these exciting
developments by some of those involved (cf. \cite{carter},
\cite{clarke:ellis:tipler}, \cite{israel:1989}, \cite{thorne:1994}) and
I shall not try to repeat any of it here.  
Naturally, any research into a theory as complicated as general
relativity starts from what could be called its `fringes', defined
by situations close to Newtonian ones, by static or
stationary model situations,
by configurations with other
symmetry or other simplifying features, etc. 
While such studies, combined with perturbative
calculations, have led to impressive results
and far reaching extrapolations, the end of the story certainly still needs
be told. Moreover, problems which may be considered by the early pioneers as
having been settled a long time ago, may require new considerations in the
darkness of new questions.

The problem of
cosmic censorship, questions about the strength of the singularity,
and the more specific but not unrelated questions raised by the
BKL conjecture and the Penrose proposal represented the {\it guiding
problems} for much of the subsequent research on the global evolution
problem.

We may well only have scratched the surface of the domain of
highly dynamical and strong field configurations and it is far from
clear that we have exhausted the content of the theory. The general
relativistic phenomena mentioned above had been predicted as
theoretical consequences of the theory before they have been confirmed
by carefully directed observations. Why should we have reached the 
end of it ? With our restricted theoretical and observational access
to the domains alluded to above we may be missing out fundamental
facts and unexpected phenomena. The results by Choptuik
(\cite{choptuik}) on the phenomena occuring at the threshold of black
hole formation may just give us a glimpse at things to come.

The coalescence of black holes is expected to represent one of the
strongest sources of gravitational radiation. Even in the clean cut
pure vacuum situation still little can be said about this process
under general assumptions, though reliable quantitative results are
needed for analysing the data of the gravitational wave experiments. 

In recent years huge strides have been taken towards the goal of
controlling the solutions to Einstein's field equations on large scale
space-time domains. The results available now go, in any sense of the
word, far beyond what was known around 1980. Nevertheless, {\it
getting qualitative and quantitative (resp. analytical, theoretical,
and numerical) control on the long time evolution of gravitational
fields under general assumptions is still the most important open
problem in classical general relativity}.

\vspace{.3cm}

\section{The exploration of the solution manifold}

One of the main problems in controlling the behaviour of solutions
to Einstein's field equations is posed by the
Ricci-operator $R_{\nu \rho}[g]$ on the left hand side of (\ref{einst}). 
In general coordinates $(x^{\lambda})_{\lambda = 0, \ldots, 3}$ 
the unknown metric $g$ is given 
by a symmetric $4 \times 4$
matrix $(g_{\mu \nu})_{\mu, \nu = 0, \ldots, 3}$ whose entries are
functions of the coordinates. In terms of these unknowns the
Ricci-operator reads  
{\footnotesize
\begin{equation}
\label{ric}
R_{\nu \rho}[g] =
\frac{1}{2}\,\sum_{\delta, \eta = 0}^{3}\,
\left\{
\frac{\partial}{\partial x^{\delta}}\,
\left(g^{\delta \eta}
\,\left[
- \frac{\partial}{\partial x^{\eta}}\,g_{\rho \nu} 
+ \frac{\partial}{\partial x^{\rho}}\,g_{\nu \eta}
+ \frac{\partial}{\partial x^{\nu}}\,g_{\rho \eta}
\right]\right)
- \frac{\partial}{\partial x^{\rho}}\,
\left(g^{\delta \eta}\,
\frac{\partial}{\partial x^{\nu}}\,g_{\delta \eta}
\right)
\right\}
\end{equation}
\[
+ \frac{1}{4}\sum_{\lambda, \delta, \eta, \pi =
0}^{3}\,\left\{
g^{\lambda \pi}
\left(
\frac{\partial}{\partial x^{\lambda}}\,g_{\delta \pi}
+ 
\frac{\partial}{\partial x^{\delta}}\,g_{\lambda \pi}
- 
\frac{\partial}{\partial x^{\pi}}\,g_{\lambda \delta} 
\right)
g^{\delta \eta}
\left(
\frac{\partial}{\partial x^{\rho}}\,g_{\nu \eta}
+ 
\frac{\partial}{\partial x^{\nu}}\,g_{\rho \eta}
- 
\frac{\partial}{\partial x^{\eta}}\,g_{\rho \nu} 
\right)
\right.\quad
\]
\[
\quad\quad\quad\quad\quad\quad
\left.
- g^{\lambda \eta}
\left(
\frac{\partial}{\partial x^{\rho}}\,g_{\delta \eta}
+ 
\frac{\partial}{\partial x^{\delta}}\,g_{\rho \eta}
- 
\frac{\partial}{\partial x^{\eta}}\,g_{\rho \delta} 
\right)
g^{\delta \pi}
\left(
\frac{\partial}{\partial x^{\lambda}}\,g_{\nu \pi}
+ 
\frac{\partial}{\partial x^{\nu}}\,g_{\lambda \pi}
- 
\frac{\partial}{\partial x^{\pi}}\,g_{\lambda \nu} 
\right)
\right \},
\]
}
where we use the coefficients $g^{\delta \eta}$ of the matrix 
$(g^{\delta \eta})$ inverse to $(g_{\mu \nu})$, which are  of the
form 
$g^{\delta \eta} = \left(\det(g_{\mu \nu})\right)^{-1} p^{\delta\eta}$ 
with polynomials $p^{\delta\eta}$ of degree $3$ in the 
$g_{\mu \nu}$.

In the case of a vacuum field with vanishing cosmological
constant
the matter
fields and the energy-momentum tensor vanish and the
various difficulties arising from matter models and matter
equation are not present. Equations (\ref{einst}) then simplifies to the
{\it vacuum field equations} 
\begin{equation}
\label{vaceinst}
R_{\nu \rho}[g] = 0,
\end{equation}
and the operator (\ref{ric}) is all one needs to consider. But
even somebody working in PDE theory for many years may  just notice
that (\ref{vaceinst}) is a equation of second order which is linear in
the second derivatives of $g_{\mu \nu}$ (i.e. quasi-linear), quadratic
in the derivatives of $g_{\mu \nu}$ and rational in the unknown $g_{\mu
\nu}$ but then may feel lost. Working out the content of the field
equations without taking into account their geometric background is a
hopeless task. Much of the richness of the system and important paths
towards analysing the structure of its solutions will remain hidden.

Usually one has to deal with in addition to  equations (\ref{einst}) suitable
matter models and the resultant {\it matter equations}. For physical
reasons it is clearly most important to understand the behaviour of the
resulting coupled systems. Matter equations import, however, their own
specific difficulties (cf. \cite{rendall: 2002}).  It is only because we do
not want to be distracted by these individual properties that almost nothing
will be said about results on solutions with matter fields.  For several
reasons the cosmological constant attracted increasing interest in recent
years and there are available quite a few results on the corresponding
solutions. Nevertheless, only the case
$\lambda = 0$ will be considered in the following.

Under {\it general assumptions} (no symmetries, algebraically
non-restricted curvature tensors, no approximation requirements like
low speeds etc.) qualitative results on the solutions to
(\ref{vaceinst}), in particular on large space-time domains, can only be
obtained by  applying geometric and abstract PDE methods to the
analysis of the field equations, subject to various suitably chosen boundary
conditions. 
Before formulating boundary value problems and proving anything, one
needs ideas about fruitful problems and a hunch of what
might be provable. Physical questions usually lead to
deep mathematical problems whose analysis requires a
considerable amount of ingenuity and the invention of new methods. 
But in the end they often allow for natural answers which
illustrate the remarkable coherence of the theory.

In the last 90 years there has been developed a host of methods to
deduce information and heuristic results from the equations. The study
of certain explicit solutions exhibiting unexpected features,  formal
expansion type analyses of the solutions based on various 
representations of the field equations, studies of perturbations away
from well-understood situations, and the use of topological,
differential geometric, and PDE methods gave rise to far reaching
generalizations.

In recent years numerical calculations have proven a
powerful extension of the arsenal. A remarkable example is provided by
the countable family of smooth, static, spherically symmetric
solutions to the Einstein-Yang-Mills equations discovered by numerical
methods by  Bartnik and McKinnon (\cite{bartnik:mckinnon}). These
solutions certainly would not have been discovered by purely
analytical methods up to this day (cf.
\cite{smoller:wasserman:yau:mcleod} for the complications of the
analytic proof).  

A closer interaction between the analytical and numerical relativists
appears to have a huge potential for further progress in the
domain of relativity we are discussing here. In the following I
shall try to point out some domains where such collaborations have been
successful and also questions where a collaboration
would be desirable and fruitful or even necessary. For a discussion of
questions more specific to numerical calculations I refer to the 
article \cite{lehner:reula} and the literature given there. 

\vspace{.5cm}

The abstract analysis of Einstein's field equations developed slowly.
That these equations themselves satisfy the
requirement of local causality was shown by
Stellmacher (\cite{stellmacher:1938a}) only in 1938. The first general local
existence proof for solutions to Einstein's equations was given in
1952 by Choquet-Bruhat (\cite{foures-bruhat}). 
Until 1980 only existence locally in time was
considered  (cf. \cite{chouqet-bruhat:york}, \cite{fischer:marsden:1979} for
surveys). Choquet-Bruhat and Geroch obtained, however, an important 
uniqueness result, which also sheds light on the dangers of using the word
`global' (\cite{choquet-bruhat:geroch}). It states that
an initial data set for Einstein's vacuum field equations determines a
{\it maximal}, globally hyperbolic time evolution {\it uniquely up to
isometries}.

Here a space-time (not necessarily solving any
equations) $(M, g)$ is called {\it globally hyperbolic} if it contains
a space-like hypersurface $S$ such that each causal curve which cannot
be extended as such meets $S$ in precisely one point. In that case
$S$ is called a {\it  Cauchy hypersurface}. 
Let $(M, g)$ be a globally hyperbolic space-time and $S$ a Cauchy
hypersurface in $M$. Cauchy data for the linear wave equation $\Box_g \phi =
0$ on $S$ then determine a solution which exists everywhere on $M$ and which
is unique. It will be shown below that the vacuum field equations can be
understood as a system of (non-linear) wave equations. This makes the
global uniqueness result plausible, but there are subtleties.

There are known maximal globally hyperbolic solutions to
Einstein's equations which can locally be extended (as solutions) but the
extension will not be unique. The hypersurface across which the
extension takes place is called a {\it Cauchy horizon}. 
The presence of a Cauchy horizon signals a {\it loss of uniquess},
which is certainly something to worry about if the field
equations are to predict the future. There arise delicate points
here. So far we assumed everything to be smooth. Should we take the
extension seriously if it can only be performed with low
smoothness ? Which smoothness requirements will still make sense from
the PDE point of view ? When do extensions still admit reasonable
physical interpretations ? Where is the dividing line between
`extension of low smoothness' and `weak singularity' ?

\vspace{.3cm}

One reason for the slow start of the general, abstract analysis of
Einstein's equations is certainly the fact that already in the vacuum case
the investigation of boundary problems for Einstein's field equations
requires the study of four different differential systems:

$-$ the {\it system of constraints},

$-$ the {\it gauge system},

$-$ the {\it main evolution system} or {\it reduced system},

$-$ the {\it subsidiary system}. 

The need for this is due to the
diffeomorphism invariance of Einstein's field equations. 
In an initial value problem equation (\ref{vaceinst})
cannot have a unique solution in the sense of PDE theory. There exists
a large set of diffeomorphisms which leave the metric  invariant up to
second order on the given initial time-slice. Therefore one has to take
special measures to reduce a problem for Einstein's equations to a
standard PDE problem. Moreover, without further assumptions the
operator
(\ref{ric}) is not of the hyperbolic type which one might 
expect for an equation required to respect local causality. 

The systems above are not independent of each other. While some are
fixed by the given boundary problem, there is a large freedom in the
choices of others and changes of one of them ususally entails
considerable changes in the others. A deeper understanding of their
meaning, freedom, and interaction is required each time new problems
are analysed. This will be illustrated below in the case of the
subsidiary system. While it is of marginal interest in the
analytical arguments, it seems to acquires an important
role in the numerical construction of space-times. 

In the following the Cauchy problem will be considered most often.
Of the many boundary problems studied for Einstein's equations,
this is the most important but not the
only relevant one. Under specific circumstances or as auxiliary
problems other boundary problems may be equally important. 
To illustrate some of the questions which need to be considered in the
context of existence theorems or in the numerical calculation of
space-times, I begin with some remarks on the systems mentioned
above and point out some old and new results.

\subsection{The constraint equations}
\label{constraints}

In the Cauchy  problem for Einstein's field equations a
prospective solution 
$(M, g)$ of equation (\ref{vaceinst}) is characterized in terms of data
prescribed on a $3$-manifold $S$ which is envisioned as being embedded
as a space-like hypersurface into $M$. Since the
initial data determine the solution near the initial hypersurface
uniquely, they contain the basic information on the solution. The
construction, detailed understanding and interpretation of initial
data is thus an important part of the Cauchy problem. Of the many
results available now only a few can be discussed here and we refer to
the recent survey
\cite{bartnik:isenberg} for more results and details of the methods
only indicated in the following.

It turns out that the data which need to be prescribed in a Cauchy
problem for the vacuum field equations (\ref{vaceinst})
are given by symmetric tensor fields $h_{ab}$ and
$\chi_{ab}$ which by the embedding of $S$ into $M$ will acquire the
meaning of the Riemannian metric and the second fundamental form
induced on
$S$ by the prospective solution $g$. 
As a consequence of the covariance of the field equations the data need
to satisfy the {\it vacuum constraint equations}
\begin{equation}
\label{constr}
R[h] - \chi_{ab}\,\chi^{ab} + (\chi_a\,^a)^2 = 0,
\quad \quad
D_c\,\chi_a\,^c - D_a\,\chi_c\,^c = 0,
\end{equation}
where $D$ and $R[h]$ denote the covariant derivative and the
Ricci-scalar of the metric $h_{ab}$ and the latter is used to move
indices. These four quasi-linear equations form an 
underdetermined elliptic system for the twelve components of $h_{ab}$ and
$\chi_{ab}$. If one takes into account the freedom to perform
transformations of the three coordinates and considers $S$ as being
determined in $M$ essentially by the {\it mean extrinsic curvature}
$\psi = \frac{1}{3}\,\chi_a\,^a$, the rough function counting gives two
degrees of freedom for the gravitational field.

The analysis of these equations depends on conditions which are not or
only partially controlled by Einstein's equations such as the topology
of $S$ or the fall-off behaviour at infinities. Of particular importance
are the cases where $S$ is {\it compact} with arbitrary topology, where
$(S, h_{ab})$ is {\it asymptotically flat} or {\it asymptotically
euclidean} (corresponding to  a hypersurface extending to space-like
infinity where the metric $h_{ab}$ approaches an euclidean metric),
or where $(S, h_{ab})$ is {\it hyperboloidal} (corresponding to a
hypersurface extending to null infinity, like a space-like unit
hyperbola on Minkowski space). In each of the non-compact cases one
may consider  $k \ge 1$ {\it asymptotic ends}, so that for some compact
subset
$K$ of $S$ the manifold $S \setminus K$ has $k$ components each of
which is diffeomorphic to $\mathbb{R}^3 \setminus B$ where $B$ is a closed
ball in
$\mathbb{R}^3$. 

In these cases the construction of solutions to the constraints is
well understood if the mean extrinsic curvature $\psi$ is assumed to be
constant ($\psi = 0$ in the asymptotically flat, $\psi = const.
\neq 0$ in the hyperboloidal case). Following a suggestion by
Lichnerowicz and using the behaviour of the  equations under conformal
rescalings of the metric, one finds that the metric can be chosen in
the form 
$h_{ab} = \phi^4\,\bar{h}_{ab}$ with some positive scalar function
$\phi$ which is to be determined by solving some equation and a metric
$\bar{h}_{ab}$ which is to be prescribed on $S$. If the symmetric trace-free
tensor 
$\bar{\chi}_{ab}$ then satisfies with respect to the metric
$\bar{h}_{ab}$ the equation $\bar{D}_c\,\bar{\chi}_a\,^c = 0$
the fields $h_{ab}$ and 
$\chi_{ab} = \phi^{-2}\bar{\chi}_{ab} + \psi\,h_{ab}$ satisfy the
vacuum constraints. This method, referred to as the {\it conformal
method}, reduces the problem of solving the constraints
to a linear elliptic system to obtain $\bar{\chi}_{ab}$
and a decoupled semi-linear elliptic scalar
equation for the conformal factor $\phi$, called the {\it Lichnerowicz
equation}. 

The solvability conditions for the latter in the compact
and to some extent in the asymptotically flat case came along with the
complete clarification of the Yamabe problem (a long standing
mathematical problem, the final step of which was taken by Schoen who,
remarkably, used ideas introduced by general relativity  (\cite{schoen:1984},
cf. also \cite{lee:parker})). The criterion is given in terms of the
sign of the {\it Yamabe number}, an invariant of the conformal structure
defined by $(S, \bar{h}_{ab})$. 
The general solvability condition in the
asymptotically flat case has been given only recently by Maxwell
(\cite{maxwell:2004}). It requires the (suitably generalized) Yamabe number
to be positive. In the hyperboloidal case solvability conditions
do not arise. 

The conformal method provides large classes of solutions to the
constraint equations. In the case of non-constant mean extrinsic
curvature $\psi$ the conformal method does not lead, however, to a
decoupling of the equations and the resultig simplifications. 
The existence of solutions to the constraints can still only been shown 
under severe conditions on $\psi$. 
The resulting restriction on the class of space-times which can be
constructed from such data may be quite serious. 
Chru\'sciel, Isenberg, and Pollack have shown the existence of 
vacuum space-times  which do not admit maximal (case
$\psi = 0$) slices (\cite{chrusciel:isenberg:pollack:2004}).
Moreoever, time slices with $\psi \neq const.$ are encountered quite
often in discussions of the evolution problem and the conformal method
can not be applied to discuss the data induced on such slices.
If such slicings are used in numerical studies it appears difficult to
replace such data by improved data which are close to the given ones
and satisfy the constraints with higher accuracy. 

\vspace{.3cm}

We have restricted our discussion to the vacuum case for
convenience only, there do exist methods to provide data for Einstein's
equations coupled to various matter fields. In those cases the data
often have a direct {\it physical interpretation}. If the data
comprise, for instance, a  ball of perfect fluid with a vacuum
exterior (cf. \cite{dain:nagy} for a detailed discussion of this
situation and its subtleties) we have a fairly clear idea about their
meaning, though the large freedom to dispose of the exterior vacuum
field still raises questions (cf. section \ref{globasflat}). 

In the pure vacuum case the physical
meaning of the data is not so obvious and can in general hardly be
assessed without analysing their evolution in time
(cf. section \ref{globasflat}).
Some interpretation if obtained for vacuum data with special
properties. The singularity theorems and the cosmic censorship
principle suggest that asymptotically flat data containing a trapped
surface, an embedded surface $\Sigma$ which is characterized by
certain convergence properties of the out- and ingoing family
of light rays orthogonal to $\Sigma$, develop into space-times containing
event horizons and black holes. Evolving such data, which arise, for
instance, if several asymptotic ends are present, is thus the usual method
to model black holes and their coalescence. 

In numerical calculations, it may be advantageous to start from
asymptotically flat initial data on a manifold $S$ with an inner
boundary $\Sigma = \partial S$ which represents a trapped surface.
Depending on the precise conditions to be achieved on $\Sigma$,
the conformal method leads in this case to various overdetermined
elliptic boundary value problems. Recently the nature of such
problems has been analysed and existence theorems have been proven
by Dain (\cite{dain:2004}), Dain, Jaramillo, Krishnan
(\cite{dain:jaramillo:krishnan}), and Maxwell
(\cite{maxwell:2004}). The initial data sets so obtained comprise
exterior data which arise  from non-trivial topologies as considered
above as well as exterior data extending to data on $S = \mathbb{R}^3$ as
described in section \ref{globasflat}.

If asymptotically flat or hyperboloidal data are to be calculated
numerically without imposing cut-offs at artificial finite boundaries,
there arises the problem that the data can develop logarithmic
sigularities at the asymptotic ends. In the asymptotically flat case
conditions on the `free data' under which logarithmic sigularities do
or do not occur at space-like infinity have been discussed by Dain and
Friedrich (\cite{dain:friedrich}). The analogous question for
hyperboloidal data has been discussed 
by Andersson, Chru\'sciel, and Friedrich (\cite{friedrich:ACF})
and in great generality by Andersson and Chru\'sciel
(\cite{andersson:chrusciel:as},
\cite{andersson:chrusciel:ph}).

\vspace{.3cm}
 
Data with quite unexpected properties have been obtained
recently by {\it gluing techniques}. A particularly remarkable idea
has been introduced by Corvino in \cite{corvino}.
The underlying method to exploit the underdeterminedness of the
constraint equations to obtain
{\it smooth, localized deformations of solutions to the constraints}
has been extended by Chru\'sciel and Delay 
(\cite{chrusciel:delay:2003}) and Corvino and
Schoen (\cite{corvino:schoen}). It provides a deeper understanding of
the  {\it constraint map} $\Phi$, which maps the fields $h_{ab}$,
$\chi_{ab}$  onto the expressions on the left hand sides of equations
(\ref{constr}) and it allows one to construct solutions to the
constraints which are not accessible by the conformal method. 
In particular, it enables one to deform given asymptotically flat
solutions to the vacuum constraints outside a given compact set to
solutions which are {\it exactly} static or stationary {\it in a
neighbourhood of space-like infinity} or to solutions which are {\it
asymptotically} static or stationary {\it at space-like infinity up to a
given order or at all orders}.

As discussed below, the surprising freedom in modifying data in their
asymptotic domain sheds a new light on  the Penrose proposal and it
raises subtle conceptional questions about the calculation of wave
forms characterizing isolated self-gravitating systems.

\vspace{.3cm}

Quite different aspects of initial data sets
are addressed in the investigations of {\it
Penrose inequalities}. 
With any asymptotically flat initial data set can be associated a
certain invariant called the {\it total mass} or ADM-mass (\cite{bartnik}).
It is obtained by performing a certain integral over a large sphere and
taking its limit when this sphere is pushed to infinity so as
to encompass the whole manifold $S$. In view of the weak
conditions imposed on the initial data by the constraints it is quite 
remarkable that this mass could be shown to be non-negative and
to be zero only for flat data.  The Penrose inequalities may be
considered as extensions of this positive mass theorem by Schoen and
Yau (\cite{schoen:yau}) and Witten (\cite{witten}).
In the case of a space-like hypersurface
$S$ embedded in a space-time with event horizons so that the induced
initial data set is asymptotically flat and $S$ intersects the
event horizon in a $2$-surface $\Sigma$, these inequalities are
expected to give  a lower bound for the total mass of the initial data
set in terms of the square root of the area of $\Sigma$.
Remarkably, under the condition $R[h] \ge 0$ on the Ricci scalar and
certain assumptions which simplify the identification of the surface
$\Sigma$, Penrose inequalities have been proven by Huisken and
Ilmanen (\cite{huisken:ilmanen}) and by Bray (\cite{bray:2001})
(cf. \cite{bray:chrusciel} for a survey).

A related class of problems is that of associating with an extended
but finite space-time domain a notion of energy or energy-momentum.
While some of the suggestions considered here played an important role
in the above proofs of the Penrose inequalities, there does not exist
a general agreement on the ultimative notions of quasi-local
energy-momentum and other quasi-local quantities (cf.
\cite{szabados} for a detailed survey).  

It is interesting that Penrose arrives at the type of inequality named
after him by invoking the $4$-dimensional space-time picture and using
a chain of arguments each of which raises questions itself
(\cite{penrose:inequ}). He assumes in particular a version of weak
cosmic censorship and makes use of the idea that after developing
(something which is to become) an event horizon the space-time will 
settle down `in some appropriate but as yet ill-defined sense' to
become a Kerr black hole. This idea is supported 
by the results on the black hole equilibrium problem 
(cf. \cite{carter}) but a proof would require control on the long time
evolution and estimates which describe in detail how the corresponding
member of the Kerr family will be approached.

Once the Penrose inequality can be derived by relying only on
properties of initial data sets, one may ask whether 
the argument could be turned around and the inequalities or the
techniques underlying their proofs could be used to obtain estimates
to control the evolution of black holes. This will, however, almost
certainly require a proof of the appropriate Penrose inequalities under
sufficiently general assumptions (including perhaps hyperboloidal
data). As discussed above, general mean extrinsic
curvatures create difficulties in solving the constraints.  They
also create difficulties in the present context. The first of
equations (\ref{constr}) shows that the condition
$R[h] \ge 0$ may be violated in the case of an unrestricted mean
extrinsic curvature. New methods may be needed to obtain the
inequality in such cases.

\subsection{The gauge system}
\label{gauge}

To apply PDE techniques to the local evolution problem, one has to 
impose {\it gauge conditions}, restrictions on the freedom to
perform diffeomorphisms or coordinate transformations. In the numerical
calculation of space-times a number of unsolved questions are
related to the gauge problem. To illustrate the general
argument and related problems without being too vague, I indicate one
specific {\it reduction procedure} by which the initial value
problem for Einstein's field equations is cast into a Cauchy problem
for a hyperbolic system. The one chosen here yields the most concise
expressions.

In the given local coordinates $x^{\mu}$ on $M$ the expression (\ref{ric})
can be rewritten in the form
\begin{equation}
\label{einlanc}
R_{\mu \nu} = - \frac{1}{2}\,g^{\lambda \rho}\,g_{\mu \nu,\lambda \rho} +
\nabla_{(\mu}\Gamma_{\nu)}
+ \Gamma_{\lambda}\,^{\eta}\,_\mu\,g_{\eta \delta}
\,g^{\lambda \rho}\,\Gamma_{\rho}\,^{\delta}\,_\nu
+ 2\,\Gamma_{\delta}\,^{\lambda}\,_{\eta}\,
g^{\delta \rho}\,g_{\lambda(\mu}\,\Gamma_{\nu)}\,^{\eta}\,_{\rho}.
\end{equation}
Here the comma indicates partial derivatives, the
$\Gamma_{\nu}\,^{\mu}\,_{\eta} = 
1/2\,g^{\mu \lambda}\,(g_{\lambda \eta,\nu}
+ g_{\nu \lambda,\eta} - g_{\nu \eta,\lambda})$ are the Christoffel
symbols, $\nabla$ is the covariant derivative operator of $g_{\mu\nu}$,
and the summation rule applies. 
The $\Gamma^{\mu}$ denote the contracted Christoffel symbols
$\Gamma^{\mu} = g^{\nu \eta}\,\Gamma_{\nu}\,^{\mu}\,_{\eta}$
which (together with the functions 
$F_{\nu}  = g_{\nu \mu}\,F^{\mu}$ considered in the following) are being
formally treated as if they defined a vector field (which, of course, they do
not).  Thus $\Gamma_{\nu} = g_{\nu \mu}\,\Gamma^{\mu}$ and $
\nabla_{\mu}\Gamma_{\nu} =
\partial_{\mu}\Gamma_{\nu} -
\Gamma_{\mu}\,^{\lambda}\,_{\nu}\,\Gamma_{\lambda}$. 

The form (\ref{einlanc}) emphasizes the first term on the right hand
side of (\ref{ric}) which is obtained by applying to the unknown
$g_{\mu \nu}$ a wave operator, a type of differential operator for
which a good theory is available. The following three terms of second
order in (\ref{ric}) prevent the direct application of PDE results.
They are hidden in the representation (\ref{einlanc}) in the second term
on the right hand side. It turns out that the apparent difficulties
dissolve once the role of the contracted Christofffel symbols
$\Gamma^{\nu}$ is recognized (\cite{friedrich:1hyp red}).

Let $(M, g)$ denote some Lorentz manifold and let $S = \{t = 0\}$, with
some coordinate function $t$, be some space-like hypersurface of it.
Consider a map $\mathbb{R}^4 \ni x^{\lambda} \rightarrow
F^{\mu}(x^{\lambda}) \in \mathbb{R}^4$. Ignoring subtleties arising from
differentiability questions, everything is assumed to be smooth. 
Local Cauchy data $x^{\lambda}$, $\partial_t
x^{\lambda}$ on $S$ determine a unique local solution to the Cauchy
problem for the semi-linear system of wave equations
\begin{equation}
\label{gaugeequ}
\Box_g\,x^{\mu} = -  F^{\mu}(x^{\lambda}),
\end{equation}
where $\Box_g = \nabla_{\nu'}\,\nabla^{\nu'}$ denotes the scalar wave
operator defined by
$g$. If the $dx^{\mu}$ are chosen initially to be pointwise
linearly independent, the solution provides a local coordinate system
$x^{\mu}$. In terms of these coordinates the system above simply takes the
form
\begin{equation}
\label{gaugeequad}
\Gamma^{\mu}(x^{\lambda}) = F^{\mu}(x^{\lambda}).
\end{equation}
As a consequence

$-$ by a suitable choice of coordinates the contracted Christoffel
symbols\\ 
\hspace*{.93cm}can locally be made to agree with any prescribed set of
functions $F^{\mu}$, 

$-$ in turn, these {\it gauge source functions} and the initial data
determine\\ 
\hspace*{.93cm}the coordinates uniquely, 

$-$ for a given metric $g$ any coordinate system is characterized by
suitable\\ 
\hspace*{.93cm}gauge source functions and initial data.

This suggests replacing the functions
$\Gamma^{\nu}$ in (\ref{einlanc}) by freely chosen gauge source functions
$F^{\nu}$. The vacuum field equations then take the
form 
\begin{equation}
\label{redequ}
0 = R^F_{\mu\nu} \equiv 
- \frac{1}{2}\,g^{\lambda \rho}\,g_{\mu \nu,\lambda \rho}
+
\nabla_{(\mu}\,F_{\nu)}
+ \Gamma_{\lambda}\,^{\eta}\,_\mu\,g_{\eta \delta}
\,g^{\lambda \rho}\,\Gamma_{\rho}\,^{\delta}\,_\nu
+ 2\,\Gamma_{\delta}\,^{\lambda}\,_{\eta}\,
g^{\delta \rho}\,g_{\lambda(\mu}\,\Gamma_{\nu)}\,^{\eta}\,_{\rho},
\end{equation}
of a system of quasi-linear wave equations for the $g_{\mu \nu}$,
which represents the {\it main evolution or reduced system} of our
procedure.  For this system the local Cauchy problem with appropriate
data on a space-like hypersurface $S$ is well posed, which means
that there can be shown the existence and the uniqueness of solutions
and their stable dependence on the initial data (cf.
\cite{friedrich:rendall} for a detailed discussion). A few
interesting observations to be made about this procedure. 

The metric coefficients $g_{\mu \nu}$ obtained as solution to 
(\ref{redequ}) are given in terms of coordinates $x^{\mu}$ which are
determined implicitly by the gauge source functions, the initial
data, and equation (\ref{redequ}). Do these coordinates really satisfy  
the {\it gauge system} (\ref{gaugeequ}) resp. its
implicit coordinate expression (\ref{gaugeequad}) ? For the moment we
assume this to be the case and consider this question again in section
(\ref{subsequ}).

The domain on which the $x^{\mu}$ form a good coordinate system
depends on the initial data, the gauge source function, and on the 
solution $g$ itself. Since information on $g$ is only acquired by
solving (\ref{redequ}), little can be said a priori on the domain of
existence of the coordinates. In this respect there is in general no
difference between {\it harmonic coordinates} (more appropriately called
now {\it wave coordinates}), characterized by
$F^{\mu} = 0$, and coordinates defined by other gauge source
functions. Without special precautions nothing prevents the $dx^{\mu}$ to
become linearly dependent, the slices $x^0 = t = const.$ with $\{t = 0\} = S$
to turn time-like, the coordinates to develop an undesirable
asymptotic behaviour, etc. If a coordinate system turns bad, one will
have to construct further coordinates and may end up with a collection
of overlapping coordinate patches which serve to define the manifold
$M$ underlying the solution space-time.

In practice, in particular in numerical calculations of space-times,
one would like to avoid such situations and try to find coordinates
which cover the entire solution.  An interesting way of
using gauge source functions to control the evolution of the slicing
in a numerical code has recently been put forward by Pretorius in his
work\footnote{In preparation} following up \cite{pretorius}. It
remains to be seen whether the method will allow him to steer the
slicing unscathed through all the dangers of a long time evolution.  

Lindblad and Rodnianski discuss the small data, global
existence for the Einstein vacuum equations in wave coordinates and it
turns out that the coordinates can be arranged to cover the complete
solution space-time (\cite{lindblad:rodnianski}). These solutions have
an asymptotic structure qualitatively similar to that of Minkowski
space.  If the data were slightly changed one would expect the
coordinates to be still well behaved by stability considerations.
Such considerations become quite delicate in global problems, however,
and unexpected things may happen when the data are increased to admit
the development of singularities and black holes.
In any case the behaviour of the coordinates has to be
controlled in the context of the evolution problem, in which little is
known about the metric a priori.

In \cite{friedrich:cg on vac} have been discussed coordinates which
cover the entire Schwarschild-Kruskal space-time up to the
singularity and even extend smoothly through the null infinities.
Again, it is unclear how they will behave if the underlying solution
space-time is perturbed. This general class of coordinates is based on
certain geometrically  distinguished curves called {\it conformal
geodesics} and the defining equations of these curves contain elements which
might allow one to control the behaviour of the coordinates over long time
intervals. While these coordinates can be characterized {\it in
principle} in terms of the gauge source function considered above,
there is no way of identifying these functions without knowing the
solution $g$. Therefore one needs a different reduction procedure to
incorporate these coordinates into hyperbolic evolution equations
(\cite{friedrich:AdS}). In this reduction the coordinates are {\it
characterized by explicit conditions on the unknowns} and one might
hope that the question which we left open above does not arise. It
comes back as the question of certain
constraints being satisfied during the evolution. 

The last example shows that it is not easy to identify good gauge
source functions. It shows also that there may be good
reasons for analysing reductions different from the one indicated
above. The role of the gauge source functions may be assumed then by
quite different quantities.  Motivated by problems in numerical
relativity there has been considered a large variety of different
reductions in recent years, based on different representations of the
field equations, different unknowns, and different types of gauge
conditions. Many of these reductions lead to  hyperbolic main
evolution systems.

\subsection{The main evolution system}
\label{main}

The main evolution system is clearly most important for working out
local existence, uniqueness, smoothness, and more specific
properties of solutions and for calculating solutions numerically.
To avoid entering technicalities I shall only
make a few general remarks about it.

In analytical work the main goal in choosing this system is to be able
to exploit the intrinsic hyperbolicity of the equations. This does not
mean that the system needs to be hyperbolic. Some useful gauge
conditions are elliptic in nature and there have been studied for
instance hyperbolic-elliptic reduced systems by Andersson and Moncrief
(\cite{andersson:moncrief:2003}). 

In recent years there has been a tendency in the general investigation
of non-linear evolution equations to study solutions of low
smoothness (\cite{tataru}). Klainerman and Rodnianski
(\cite{klainerman:rodnianski}) study 
Einstein evolution under smoothness assumptions which are weaker than
those considered up to a few years ago (cf. \cite{hughes:kato:marsden}) 
and Maxwell (\cite{maxwell:2004b}) discusses the existence
of rough solutions to the constraints equations. If one is mainly
interested in physical phenomena, these activities may appear quite esoteric.
As indicated above,  questions about precise and low smoothness requirements
can become, however, inavoidable in the discussion of singularities and
Cauchy horizons. Moreover, trying to push to their lower limits the
smoothness requirements under which the Einstein equations still make sense
forces one to explore the specific structure of the equations much more
carefully and it is bound to lead to more precise information on the
evolution.  

Once local existence has been
treated, there may be used other methods to get control on the long
time evolution. While some hyperbolic main evolution systems imply
energy estimates involving the Bel-Robinson tensor, this tensor
may be used independent of any reduction to derive estimates on the
solution. 

In any case there needs to be erected some kind of rigid space-time
structure, a foliation by space-like or null slices or a fixed coordinate
system, relative to which estimates of the metric field are expressed and the
evolution of this structure itself can be controlled. In
the case of the global or semi-global non-linear stability results 
mentioned below one can lean back on information supplied by an explicit
reference solution, choose on it a foliation determined by some suitable
(evolution) law, and construct a foliation on the perturbed solution
governed by the appropriately perturbed law. If the coalescence of two black
holes is to be modelled under general assumptions, however, there are no
reference solutions available and one has to develop an intuition for
foliations with  long life times and good evolution properties in the
context of an existence proof or in the course of an numerical calculation.

These remarks also show that the needs of analytical and numerical work
are different. In the latter one has to rely on an explicit main
system for all times of the evolution (or at least change the system
only a finite number of times). For the choice of this system the main
requirement is the stability of the numerical evolution. 
So far this has essentially been a matter of trial and error
and there appears to be no way to translate this requirement into a precise
criterion in terms of algebraic or other properties of the system. Manifest
hyperbolicity is reasonable but apparently not sufficient for that
purpose. There is in use a class of systems, the so-called BSSN
systems (\cite{baumgarte:shapiro}, cf. also \cite{alcubierre:etal}, 
\cite{friedrich:rendall}, \cite{fritelli:reula:II} for hyperbolic
versions) some of which are not manifestly hyperbolic but nevertheless seem
to lead to stable numerical evolutions. Understanding
why this should be so or for which class of problems stability fails
for these systems is a theoretical challenge.

\subsection{The subsidiary system}
\label{subsequ}

In our discussion of the gauge system we left open the question
whether the implicit gauge condition (\ref{gaugeequad}) is preserved
under the evolution defined by (\ref{redequ}). Here is the
analytical argument. Equation (\ref{redequ}) is of the form 
\begin{equation}
\label{3redequ}
R_{\mu \nu} = \nabla_{(\mu}\,Q_{\nu)},
\end{equation}
where $Q_{\mu} = \Gamma_{\mu} - F_{\mu}$ with the $\Gamma_{\mu}$
calculated  from the solution $g_{\mu \nu}$. The requirement that 
$Q_{\mu} = 0$ may be considerd as a kind of constraint for
(\ref{3redequ}) (it is in fact related to the constraints considered in
section (\ref{constraints})).  The twice contracted Bianchi identity,
which holds for any metric, reads $\nabla^{\mu}(R_{\mu \nu} -
\frac{1}{2}\,R\,g_{\mu \nu}) = 0$. Applying this to the equation above
gives
\begin{equation}
\label{subs}
\nabla_{\mu}\,\nabla^{\mu}\,Q_{\nu} + R^{\mu}\,_{\nu}\,Q_{\mu} = 0,
\end{equation}
and thus a {\it subsidiary system} of wave equations for the quantities
$Q_{\nu}$.

It turns out that for data satisfying the constraints and the gauge
condition $Q_{\mu} = 0$ on $S$ the solution to (\ref{redequ}) 
satisfies $\partial_t Q_{\mu} = 0$ on $S$. Since we have thus vanishing 
Cauchy data for the hyperbolic system (\ref{subs}), it follows that
$Q_{\mu}$ vanishes and
$g_{\mu \nu}$ does solve (\ref{vaceinst}). This closes the
argument in the continuum model. Note that only the
homogeneity and the uniqueness property of (\ref{subs}) are being used
here. There is no further role for (\ref{subs}) in
analytical studies.

\vspace{.3cm}

The situation is quite different in the discrete model. Most numerical
calculations of solutions to Einstein's equations are being
plagued by an undesirably fast growth of  constraint violations. In
fact, many workers in the field report on seemingly unmotivated
catastrophic blow-ups of numerical calculations at a stage of the
numerical time evolution where coordinate or curvature singularities
were not to be expected. Understanding this situation requires and
deserves a major effort. In the following I shall not discuss any of
the remedies which have been suggested (cf.  
\cite{brodbeck:etal}, \cite{gundlach:etal}), or the stability analyses
of the subsidiary system, considered as a linear system on a given
background (cf. \cite{frauendiener:vogel},
\cite{yoneda:shinkai}). I would like to identify instead 
possible sources of the problem in the analytical structure of the
equations. 

Analytically, (\ref{subs}) is simply a differential identity implied by
(\ref{3redequ}). It is hard to see, however, how to devise a numerical
scheme for the second order wave equations 
(\ref{3redequ}) for which the subsidiary system (\ref{subs}), which is of
third order in the metric, could be identified as an identity.
The relations between the two systems will therefore become obscured,
and the development of the constraint violation is not easy to
analyse. 

If we observe (\ref{3redequ}) in  (\ref{subs}), the latter takes the form
\begin{equation}
\label{nonlinsubs}
\nabla_{\mu}\,\nabla^{\mu}\,Q_{\nu} 
+ Q^{\mu}\,\nabla_{(\mu}\,Q_{\nu)} = 0,
\end{equation}
of a manifestly non-linear wave equation.
In the continuum model this equation would still imply $Q_{\nu} = 0$.
In the discrete model the quantity $Q_{\nu}$ comes, however, with an
error initially and develops further errors during the evolution. The
detailed propagation properties of (\ref{nonlinsubs}) therefore become
important. If (\ref{subs}) were a linear system on a given background,
standard energy estimates would admit and cannot exclude an exponential
growth of the unknown $Q_{\mu}$ but would admit nothing faster than
that. The non-linearity of (\ref{nonlinsubs}) might induce, however, a
much faster growth of the constraint violation.

Analysing this situation is not easy. While (\ref{3redequ}) can be
studied independently, equation (\ref{nonlinsubs}) does not decouple
from (\ref{3redequ}). The latter supplies the metric defining the
background for (\ref{nonlinsubs}) and the growth of $Q_{\mu}$ 
has an effect on the evolution of the metric by (\ref{3redequ}). The
coupling of (\ref{nonlinsubs}) to (\ref{3redequ}) defines a second,
though less direct, non-linearity in the evolution of
$Q_{\mu}$. Nevertheless,  considering the background metric in
(\ref{nonlinsubs}) as given, should provide some understanding
of the propagation properties of that equation.

If one considers (\ref{nonlinsubs}) as an equation on
Minkowski space, it turns out that it is easy to find Cauchy data 
$Q_{\mu}$ and $\partial_t Q_{\mu}$ on
$\{t = 0\}$ for which the solutions develop poles after finite
coordinate times $t_* > 0$ (cf. \cite{friedrich:2005} for details). These
data can be chosen pointwise as small as one likes, though $t_* \rightarrow
\infty$ if the data approach zero. This
suggests that one may have a relatively stable numerical evolution for
a while but at a certain stage effects due to the non-linearity in
(\ref{nonlinsubs}) take over and induce a catastrophic collapse of the
calculation after some finite time. 

There are quite a few interesting
questions to be answered and the situation certainly deserves closer
analytical and numerical study.
We have discussed here the non-linearity of the subsidiary system in a
specific example. Most likely, non-linearities are
found in all subsidiary systems. To what extent their effects are
different and whether there exist preferred cases remains to be seen.

\subsection{Global studies}

In the following we want to present some global or
semi-global results on the existence of solutions to Einstein's field
equations. The field has seen a rapid development in recent years.
It will not be possible here to give due reference to all the important
contributions and we will have to make a somewhat arbitrary choice. 
In the following it will be convenient to make a distinction between
vacuum solutions which are {\it asymptotically flat} and solutions with
{\it compact time sections}. In spite of the fact that there are
important cosmological models with non-compact time slices, by
a {\it cosmological space-time} will be meant in the following always
a vacuum solution with compact time slices.

\subsubsection{Cosmological space-times}

In the context of cosmological solutions there exists a large variety
of possible assumptions on the (local) symmetries and the topologies
of  space-sections. The tendency of much of the recent work has been to
analyse the global behaviour of the solutions under strongly
simplifying assumptions and then, relaxing them step by step, to work
ones way up to develop in the process the insight and the
technical means to analyse in the end also quite general classes of
solutions.  Of the many results obtained in this program only
a few can be considered in the following.

What are the questions to be asked ? If one tries to work out
a `global' existence result by analysing a Cauchy problem for a certain
class of solutions and it can be shown that the maximal
globally hyperbolic solutions determined by the data are all
geodesically complete in the past and in the future one may still be
interested in the precise asymptotic behaviour near past and future
time-like infinity but the essential goal has been reached.

In general the situation will be more complicated. It may happen
instead that the solution will be geodesically complete in one
time direction (or not at all) but in the other direction there are
obstruction to completeness. Extreme (and, in a sense, the clearest)
possibilities which can occur are that the solution approaches a
smooth Cauchy horizon or that it develops of a curvature singularity at
which an invariant built from the curvature tensor such as the
Kretschmann scalar $R_{\mu \nu
\lambda \rho}\,R^{\mu \nu \lambda \rho}$ becomes unbounded. This may
happen globally when a family of space-like hypersurfaces
approaches the corresponding end of the evolution or it may happen only
locally. There are possible all kinds of situations in between and
combinations thereof. One would like to have a detailed description of
the respective behaviour and relate specific types of behaviour to
the `size' of the corresponding subset of the space of initial data.

Since any statement about cosmic censorship must refer to
{\it generic} classes of initial data, the analysis of solutions with
symmetries cannot give a final answer to the question of cosmic
censorship. It can, however, provide insight into the nature of the
problem to analyse the question in the class of space-times with a
given symmetry. Moreover, if the subset of solutions which do not
admit Cauchy horizons contains a dense open subset of the data
set one would be prepared to consider this as evidence for cosmic
censorship ({\it restricted cosmic censorship}). 

The `simplest' and still quite interesting case to consider is given by
space-times which are {\it spatially homogeneous} in the sense that they
admit a 3-dimensional space of Killing fields which at each point generate
the tangent space of a foliation by (compact) space-like hypersurfaces. 
In the universal covering space the (not necessarily compact) leaves of
the foliation  can then be considered as orbits of a
3-dimensional Lie group $G$ of isometries. These {\it Bianchi
space-times} have been classified in terms of the Lie algebras of
their isometry groups. 

For spatially homogeneous space-times the Einstein vacuum equations
essentially reduce to systems of ODE's. Nevertheless, their analysis
turns out quite difficult, not because of the various cases in
the Bianchi classification but because of the various phenomena
which can occur. The Bianchi classification consists of two classes.
We shall consider only one of them, in which the structure constants
of the Lie algebras can be specified in terms of a symmetric matrix
which can be assumed to be diagonalized (case A). In that case the
Einstein vacuum equations have been written by Hsu and Wainwright 
(\cite{hsu:wainwright}) as an ODE and a constraint 
for an unknown 
$u = (N_1, N_2, N_3, \Sigma_+, \Sigma_-)$
of the form
\[
\frac{d}{dt}u = f(u), \quad q(u) = 1,
\]
with $q$ a quadratic polynomial in the components of $u$.
The first three components of $u$ contain essentially the information
on the Lie algebra and the other two information on the second
fundamental form on the leaves of the foliation.

When the Lie algebra is cummutative (Bianchi I) only
$\Sigma_+$, $\Sigma_-$ do not vanish and the constraint reduces to the 
equations $\Sigma^2_+ + \Sigma^2_- = 1$, which defines the {\it Kasner
circle} parametrizing these solutions. The fields $\Sigma_+$, $\Sigma_-$
are in fact constant, i.e. the solutions are fix points of the
dynamical system above. The solutions are given explicitly by the  {\it
Kasner metrics} 
\[
ds^2 = - dt^2 + \sum_{a = 1}^3t^{2 p_a}( dx^a)^2,
\]
on $]0, \infty[ \times \mathbb{R}^3$ or, after a periodic identification, on 
$]0, \infty[ \times T^3$. The real numbers $p_a$ satisfy the Kasner
relations $\sum_{i = 1}^3p_a = 1$ and $\sum_{i = 1}^3p^2_a = 1$
and are related to the two unknowns above by 
$\Sigma_+ = \frac{3}{2}(p_2 + p_3) - 1$ and 
$\Sigma_- = \frac{\sqrt{3}}{2}(p_2 - p_3)$. 

The three points
$(-1, 0)$, $(\frac{1}{2}, \pm \frac{\sqrt{3}}{2})$ on the Kasner
circle correspond to flat solutions, which in the case without
identifications are isometric to open subsets of Minkowski
space given by the future of the intersection of two null hyperplanes.
In the non-flat case the causal geodesics are incomplete in the past
(where $t \downarrow 0$) and approach a curvature singularity there.
Note that the volume of the time slices grows monotonically
with $t$ but the growth happens locally in an anisotropic way because
the $p_a$ can not all have the same sign.  

Of particular interest is the case where the group $G$ is
$SU(2)$ (Bianchi IX). It contains a
subclass, the Taub type IX solutions, characterized by linear
conditions on the components of $u$. These solutions can also be
given explicitly  and it turns out that they admit in the past
(compact) Cauchy horizons and smooth extensions (there do exist in fact
non-isometric extensions (\cite{chrusciel:isenberg:1993})) across it,
resulting in the Taub-NUT space-times. 

The remaining `generic Bianchi IX solutions' or {\it proper Mixmaster
solutions} (\cite{misner}) show an interesting behaviour. 
They have a curvature singularity which is approached by the solutions
with an oscillatory behaviour. The projections of their trajectories
into the $\,(\Sigma_+, \Sigma_-)$ - plane moves into the Kasner circle
and stays there, approaching subsequently different points on the
circle (cf. \cite{berger:garfinkle:strasser} and \cite{rendall: 1997},
\cite{ringstrom:2000} for numerical and analytical studies).
The detailed studies of the Bianchi models show that curvature
blow-up and the non-existence of
Cauchy horizons in the contracting direction is a feature of generic
Bianchi space-times, which shows
that retricted strong cosmic censorship holds in this class of
models (\cite{ringstrom:2000}, cf. also \cite{chrusciel:rendall}).  

While being interesting in themself, the results on the Mixmaster
solutions are expected to be of a much wider significance. The
BKL conjecture mentioned in section
\ref{radsingbh} suggests that the behaviour of these solutions
near the singularity provides a model for the local behaviour of
general cosmological solutions in the neighbourhood of singularities. 

\vspace{.5cm}

Moncrief initiated with the article \cite{moncrief:1981} the global study of
a class of vacuum solutions with two commuting space-like Killing fields
and compact space sections called {\it Gowdy space-times}. With the
spatial topology being that of the $3$-torus $T^3$ the metric can be written
in the form
\[
g = \sqrt{\tau\,e^{- \lambda}}\,(- dt^2 + dx^2)
+ t\,(e^P(dy + Q\,dz)^2 + e^{-P}\,dz^2),
\]
with $x$, $y$, $z$ each being a coordinate on the circle $S^1$, the
coordinate $t$ taking values in $\mathbb{R}_+$, and the coefficients
$\lambda$, $P$, $Q$ depending only on $t$ and $x$. 

For these metrics the Einstein vacuum equations read
\[
P_{,tt} - P_{,xx} = -\frac{1}{t}\,P_{,t} + e^P (Q^2_{,t} - Q^2_{,x}),
\]
\[
Q_{,tt} - Q_{,xx} = -\frac{1}{t}\,Q_{,t} 
- 2\,(P_{,t}\,Q_{,t} - P_{,x}\,Q_{,x}),
\]
and 
\[
\lambda_{,x} = - 2\,t\,(P_{,t} \,P_{,x} + e^{2 P} Q_{,t} \,Q_{,x}),
\]
\[
\lambda_{,t} = - t\,(P^2_{,t} + P^2_{,x} + 
e^{2 P}(Q^2_{,t} + Q^2_{,x})).
\]

If Cauchy data for $P$ and $Q$ are given for which the integral over
the circle of the expression on the right hand side
of the third equation vanishes, the discussion reduces
to the analysis of the  {\it semi-linear} system of wave equations for
the functions $P$ and
$Q$, which essentially represent the two
polarizations states of the gravitational field.

After Moncrief had given the first global existence proof for the
solutions, these solutions and the nature of their singularities were
studied by many authors. The analysis
was largely assisted by the numerical work initiated by 
Berger and Moncrief (\cite{berger:moncrief}). It showed that the
solutions tend to develop strong gradients ({\it spikes}) in the
approach towards the singularity, which brought an important aspect
into the analytical discussion. Kichenassamy and Rendall 
constructed families of solutions with singularities by Fuchsian
methods (cf. the survey \cite{rendall:2004b} and also the work by Chae and
Chru\'sciel (\cite{chae:chrusciel})). Rendall and Weaver constructed families
of Gowdy solutions with spikes from solutions without spikes, and
distinguished `true' spikes, which have a geometric
meaning, from `false' spikes (\cite{rendall:weaver}). 
Building on this and earlier work, Ringstr\"om recently showed that
for a `generic' set of initial data the corresponding solutions exhibit a
curvature blow up on dense open subsets of the singularity 
(\cite{ringstrom:2005}). Combining this result with the work by Chru\'sciel
and Lake on the occurrence Cauchy horizons in Gowdy space-times
(\cite{chrusciel:lake}), he concludes that this class of space-times
satisfies a restricted strong cosmic censorship principle.

This short discussion hardly gives credit to the many important
contributions which led to the `final' answers. A more complete picture
of these interesting developments can be obtained from the articles
\cite{andersson:2004}, \cite{andersson:van elst:uggla}, and
\cite{berger:2002}. The latter is also particularly interesting
because it highlights the remarkably successful and still
ongoing interplay between numerical and analytical studies in this
field.

The results mentioned above do not
finish the analysis of solutions with two Killing fields. In fact, the
Gowdy metrics considered here only define a `negligible' subset of the
set of all solutions with two Killing fields. 

\vspace{.5cm}

As a further step in the program solutions with only one
Killing fields should be analysed. A semi-global, non-linear
stability result in this directions is obtained by Choquet-Bruhat, who
generalizes previous work with Moncrief to show
future completeness for a class of $U(1)$-symmetric
vacuum solutions on manifolds of the form $M = \mathbb{R} \times \Sigma
\times S^1$ with Cauchy hypersurface diffeomorphic to 
 $\Sigma \times S^1$,
where $\Sigma$ is an orientable, compact surface of genus
greater than $1$ and the space-like Killing fields are assumed to be
tangent to the fibres of the fibration defined by the projection
$M \rightarrow \mathbb{R} \times \Sigma$ (\cite{choquet-bruhat:2004}).

Generalizing even further, Andersson and  Moncrief obtain a
semi-global, non-linear stability result for vacuum solutions without
imposing any symmetry conditions (\cite{andersson:moncrief:2004}) (the
solutions of \cite{choquet-bruhat:2004} are not included). Denote by
$V$ the interior of the future light cone at the origin in Minkowski
space and by $\tau$ the Minkowskian distance from the origin.
Identification of points of $V$ by the action of a suitable discrete
subgroup of the Lorentz group yields a {\it reference space-time} of
the form
\[
M = \mathbb{R} \times S, \quad g = - d\tau^2 + \tau^2\,h,
\]
where $(S, h)$ denotes a 3-dimensional, compact hyperbolic space of
sectional curvature $-1$. We assume $\partial_{\tau}$ to be future
directed. The authors consider cases where $(S, h)$
satisfies a certain {\it rigidity} condition.
Identifying $S$ with the set $\{\tau = 1\}$, the metric above 
induces on $S$ the {\it reference data} $(h_{ab}, \chi_{ab} = h_{ab})$.
It is shown in 
\cite{andersson:moncrief:2004} that vacuum data on $S$ 
sufficiently close to rigid reference data develop into solutions of
the vacuum field equations for which the causal geodesics are future
complete. The authors also desribe the asymptotic decay of their
solutions towards the reference solution.

For recent attempts to control the behaviour near the
singularity under `general' assumptions, to give precise
meaning to the BKL conjecture, and to develop tools which would allow
one to decide on its valitity, we refer to the work by
Andersson et al. (\cite{andersson:van elst:lim:uggla}) and 
Garfinkle (\cite{garfinkle:2004}). There appears to be a general
expectation that the BKL conjecture will turn out to be basically
correct.

\subsubsection{Asymptotically flat space-times}
\label{globasflat}

Asymptotically flat space-times provide the basic model 
of isolated self-gravita\-ting systems such as stars, star systems,
black holes, etc. and as such they are important for discussing many
observable general relativistic phenomena, the analysis of radiative
phenomena being at present the most important
and urgent one. The modelling of stars is clearly an important task.
Nevertheless we shall concentrate again on the vacuum case or
situations with field theoretical matter models. On the one hand,
progress in these cases will provide important insights into the
behaviour of dynamical black holes, the coalescence of black holes, the
resulting radiation fields etc., and possibly quantitative results
about the latter,   while on the other hand even these configurations still
present major challenges.  
 
Asymptotically flat space-times pose problems which do not occur in 
the cosmological context. The space-like slices are of infinite extent
and on an asymptotically flat slice 
the fall-off behaviour near space-like infinity can, in terms of coordinates
$x^{\mu}$ which realize the asymptotic flatness conditions, not be
faster then 
\[
g_{\mu \nu} - \delta_{\mu \nu} = \frac{m}{2\,|x|}\,\delta_{\mu
\nu} + o(\frac{1}{|x|})
\quad \mbox{as} \quad |x| \rightarrow \infty,
\]
unless the total mass $m$ vanishes, in which case the space-time is
flat by the positive mass theorem. This poses major problems for 
global existence proofs and, in particular, for the verification of the
Penrose proposal under general assumptions.  

The infinite extent of space-like slices also requires
additional considerations in the numerical study of
space-times, because numerical calculations need to be
performed on finite computational grids. Quite a few technical
considerations and also conceptional questions, some of which will be
indicated below, are related to that fact.

Since gravitational radiation can either escape to null infinity or
fall into a black hole, which is possibly (cf. the discussion below)
generated by the radiation itself, the completeness properties of the
space-times in the future and the structure of their time-like
infinities can vary considerably.  Already in stationary
examples such as the Schwarzschild-Kruskal, the Reissner-Norstr\"om, 
and the Kerr solution there arise extreme and quite  distinct
situations near time-like infinity, where singularities, event
horizons, null infinities, and Cauchy horizons seem to meet
in the standard causal pictures (\cite{hawking:ellis}). Hardly anything
is known about the possibilities under general assumptions. 

\vspace{.3cm}

The central question in the field is again 
whether cosmic censorship is a valid principle. Since any
precise statement about that principle will use the word generic,
the general answer can only be obtained by analytical methods.
Numerical methods are likely to play, however, a major role in
the investigation of the possibilities.    

At present the main question related to physical observations is
concerned with concepts of radiation, the precise asymptotic behaviour
at null infinity, and methods to derive quantitative results about
radiative properties. Here numerical methods are bound to play a
dominant role in the end, but before that quite a few analytical
questions will need to be answered. 

The two questions above are not independent of each other. Even if one
only seeks to calculate the radiation field, if one were not
interested in the interior of black holes and the structure of
singularities, and if one knew somehow that singularities were always
hidden behind event horizons, one could hardly ignore in the analysis
of the long time evolution of gravitational fields the
tendency of solutions to develop singularities. Furthermore, since
the location of an event horizon is not known at a finite stage of
the evolution, the domain of outer communication, comprising the far
fields, can in the usual approach to initial value problems not be
cleanly separated from the interior of a black hole. Present attempts
in numerical calculations to cut out the singularity from the
computational domain may become a delicate matter in long time
calculations. 

\vspace{.5cm}

In the case of asymptotically flat solutions the number of
interesting simplifying assumptions is much smaller than in the
cosmological context. If stationary solutions are excluded and one
insists on complete and 
regular far fields, only spherical or axial symmetry remain as 
admissible symmetries. Additional
fields need to be considered to analyse any
dynamical behaviour in a spherically symmetric setting, because spherically
symmetric vacuum solutions are, by the Birkhoff theorem, locally isometric
to patches of the Schwarzschild-Kruskal solution.  Of the large body of work
dealing with spherically symmetric situations the following three
contributions are particularly important.
 
In a remarkable series of articles Christodoulou analyses the
formation of black holes and singularities for the spherically
symmetric Einstein-scalar field equations. He gives conditions on
the data for the avoidance and for the development of singularities
respectively. He further gives conditions under which the singularity
will be hidden by an event horizon but he also finds solutions whose
singularities  can be seen by distant observers, thus showing the
occcurence of naked singularities. Finally, he shows that the
existence of naked singularities is in fact an unstable property for
the  spherically symmetric Einstein-scalar field equations, which
supports the cosmic censorship hypothesis (cf.
\cite{christodoulou:1999} and the references given there). 

In a seminal article (\cite{choptuik}) and subsequent work Choptuik
studies numerically one-parameter families of spherically symmetric
Einstein-scalar fields for which the solutions disperse for values of
the parameter below  but form black holes for values of the parameter
beyond a certain threshold value. This allows him to discover phenomena
which would have been difficult to find by purely analytic methods.
In particular, he finds self-similar
critical solutions, the vanishing of the black hole mass as the
parameter approaches its critical value, and a certain scaling of
the black hole masses with universal critical exponents. This work
initiated quite a number of further investigations involving various
different matter models and different types of non-linear
equations (cf.
\cite{gundlach} for a survey). 

Dafermos and Rodnianski study the spherically symmetric
Einstein-Maxwell scalar field equations, assuming that a regular
event horizon has formed (\cite{dafermos:rodnianski:2003}).
They give rigorous proof to Price's result (\cite{price}) that
perturbations of gravitational fields show in terms of a suitably
chosen advanced time coordinate a polynomial decay on the event
horizon near time-like infinity. Moreover, they confirm results by
Israel and Poisson (\cite{poisson:israel}) concerning the occurrence
of weakly singular Cauchy horizons (cf. also
\cite{israel:1998}, \cite {ori:1999} for the general background
and \cite{berger:2002} for numerical studies of the
interior of black holes). The significance of this result
concerning the question of strong cosmic censorship remains an open
question as long as only spherically symmetric situations are
considered (fortunately some kind of censorship is enforced already by
the presence of the event horizon).

\vspace{.5cm}

Compared with the richness of these detailed results and observed
phenomena, the study of the large scale structure of  asymptotically
flat vacuum solutions {\it without symmetries} is still in its infancy.
Compared, however, with the technical difficulties to be overcome,
quite some progress has been achieved in the last twenty years. 
All the global or semi-global results concerning
asymptotically flat solutions without symmetries which have been
obtained so far are stability results which show the existence of
geodesically complete or future complete solutions for {\it small
data}, i.e. for data which are in suitable norms close to the data of
a well underderstood reference solution, provided by Minkowski space
or parts of it.  Analogous results for comparison solutions with black
holes are not available yet.

\vspace{.5cm}

An early semi-global existence result was obtained in the article
\cite{friedrich:n-geod}, where it was shown that smooth hyperboloidal
data sufficiently close to Minkowskian hyperboloidal data develop
into solutions of the vacuum field equations which have a smooth
future complete structure at null infinity for which time-like
infinity is represented in suitable conformal extensions by a regular
point. The
analysis uses the conformal behaviour of the Einstein equations in an
explicit way and it is carried out in terms of a conformally rescaled
metric with respect to which null infinity and time-like infinity are
at a finite location. The solutions show in particular the {\it peeling
behaviour}, so that along outgoing null geodesics with affine
parameter $r$ the components of the conformal Weyl tensor satisfy in a
suitable orthonormal frame a fall-off behaviour near null infinity 
which can be expressed in terms of certain entire powers of
$r^{-1}$.

The particular initial value problem was considered to avoid in a first
step the complications at space-like infinity. While the setting was
thus intended as a preparation for a global study, it acquired in the
meantime some interest for the numerical calculation of radiation
fields based on the underlying conformal field equations
(\cite{frauendiener}, \cite{huebner:2001}, \cite{husa:tueb}). 

Soon afterwards attempts were made to use this result to construct
complete solutions. The idea was to avoid the complications at
space-like infinity by constructing Cauchy data for which the evolution
near space-like infinity could be controlled explicitly and the
existence of smooth hyperboloidal slices close to Minkowskian ones
could be shown. Cutler and Wald managed to construct non-trivial data
for the Einstein-Maxwell equations which are spherically symmetric
outside a compact set (\cite{cutler:wald}). Thus they were able to
establish the existence of geodesically complete solutions to the
Einstein-Maxwell equations with smooth and complete  structures and
non-trivial radiation fields at future and past null infinity  (cf.
\cite{friedrich:global}). More recently, using the result of Corvino
mentioned above,  Chru\`sciel and Delay showed the existence of a large
class of vacuum solutions with these properties
(\cite{chrusciel:delay:2002}). To some extent this justifies the Penrose
proposal but the fact that these solution are exactly Schwarzschild near
space-like infinity clearly leaves  space for generalizations. 

\vspace{.5cm}

The first global existence result for a general class of asymptotically
flat data was obtained by Christodoulou and Klainerman
(\cite{christ:klain}). They exploit the conformal behaviour of
the solutions only indirectly. With a considerable effort they
skillfully manage to control the behaviour of the fields near
space-like infinity.  They show that vacuum data satisfying certain
regularity conditions near space-like infinity (which include the
vanishing of the linear momentum) and suitable smallness conditions on
the initial slice develop into a unique, globally hyperbolic solution
of Einstein's vacuum field equations which is geodesically complete and
asymptotically flat in the sense that the Riemann curvature approaches
zero along any geodesic if the affine parameter tends to infinity.
The global structure of their solutions is qualitatively similar to
that of the space-times considered above, with no further conditions on the
data, however, the peeling behaviour in its usual form is not satisfied and
the  smoothness of the conformal structure at null infinity is thus weaker
than that required by the Penrose proposal. 

Using instead of a maximal foliation near space-like infinity a double
null foliation, Klainerman and Nicol\`o proof a similar though
technically simplified result (\cite{klainerman:nicolo}). Revisiting
their proof in the light of the results discussed above they give
asymptotic conditions on the data near space-like infinity which
allow them to verify the peeling behaviour
(\cite{klainerman:nicolo:II}).

More recently Lindblad and Rodnianski obtained a technically even more
simplified global existence proof by using the Einstein vacuum
equations in wave coordinates (\cite{lindblad:rodnianski}). They
avoid the difficulties near space-like infinity by starting from
Cauchy data which are exactly Schwarzschild near space-like infinity
but they plan to give a proof based on weaker assumptions in a
forthcoming article.  It is not clear yet to what extent their method
will allow them to gain precise control on the smoothness resp. peeling
behaviour of the fields near null infinity.

\vspace{.5cm}

Any analysis of the evolution of the fields near space-like infinity
will need to impose 
restrictions on the initial data to obtain
control on the evolution and on the behaviour of the fields near null
infinity. In the articles \cite{friedrich:AdS} and \cite{friedrich:i-null}
has been developed a setting which allows one to analyse under suitable
regularity conditions on the data at space-like infinity the evolution
of the fields near {\it the critical set where space-like infinity
touches the null infinities} (a notion made precise in
\cite{friedrich:i-null}) at all orders. As a result it is shown that
even for (conformal) data of maximal smoothness the solution can
develop at all orders logarithmic singularities at the critical set
and consequently on null infinity (cf. \cite{friedrich:spin-2}).
Furthermore, there is obtained for the first time a series of
analytical expressions which relates a certain class of obstructions to
the smoothness at null infinity to certain specific fall-off properties
of the initial data. While the analysis is carried out in an
algorithmic way, the complexity of certain expressions grows
so quickly with increasing order that the possible existence of a
further class of obstructions was left open.

Recently the remaining case was studied by Valiente Kroon by using an
algebraic computer program (\cite{valientekroon:2004a}). He did find
further obstructions and remarkably, up to the order to which the
calculation could be performed, there is now evidence that in
the case of time reflexion symmetric data smoothness (resp. $C^k$) at
null infinity requires the data to be {\it asymptotically static}
(resp. asymptotically static up to an order $p$ for a certain integer $p =
p(k)$ which still needs to be determined). Note that this is much weaker than
`static in a neighbourhood of space-like infinity', in which case that
smoothness at null infinity is easily shown. For more general data 
(\cite{valientekroon:2005}) the situation is not so clear yet but {\it
asymptotic stationarity} may play an important role. A more detailed
discussion of the situation is given in 
\cite{friedrich:cargese}.

\vspace{.5cm}

Since the work referred to above relies on a conformal
respresentation of the Einstein equations which only seems to be
useful in $4$ dimensions (cf. \cite{friedrich:tueb}), it may be worth
mentioning here that other  conformal field equations have been
suggested recently by Anderson which work in all {\it even} space-time
dimensions (\cite{anderson:2004}). It can be expected that many of the
results obtained in
$4$ dimensions can be generalized to all even dimensions
(\cite{anderson:chrusciel:2004}). Recent results
by Hollands and Wald (\cite{hollands:wald}) and Rendall 
(\cite{rendall:2003}) suggest, however, that conformal equations with
similar properties do not exist in odd space-time dimensions.

\vspace{.5cm}

Besides clarifying the asymptotic behaviour of gravitational fields
one of the main motivations for the work in \cite{friedrich:i-null} was
to provide a setting which would allow one to {\it calculate
numerically  entire space-times on finite grids, including their
asymptotic structure and radiation fields}. Once the numerical
evolution can been pushed past the critical set the solution will
contain hyperboloidal slices and earlier analytical and numerical
results can be applied. The numerical implementation for this program
has not been given yet. As shown by the following example, there are
interesting question which would be difficult to study by other numerical
approaches. 

Beig and O'Murchadha describe in \cite{beig:o'murachadha:1991} an
interesting construction of asymptotically flat initial data on
$S = \mathbb{R}^3$ with trapped surfaces and suggest that these surfaces are
due to concentration of gravitational radiation
(these data are not obtained by using initial hypersurfaces with
non-trivial topology but by analysing sequences $h_n$ of conformal
metrics with positive Yamabe number on $S^3$ for which the Yamabe
number approaches zero as
$n  \rightarrow \infty$). Since the data are time
reflection symmetric, the singularity to be expected in the future
must be considered as a reflection of the singularity in the past and cannot
be interpreted as being due to radiation.  In
\cite{beig:o'murachadha:1994} similar data  are constructed without the time
reflection symmetry, but this by itself does not preclude the
existence of a singularity in the past.

Dafermos has recently shown the existence of maximal
developments arising from asymptotically flat Cauchy data
for the spherically symmetric Einstein-scalar field equations, which
contain an event horizon in the future but for which all causal
geodesics are complete in the past (\cite{dafermos:2003}). One would
consider a solution to the Einstein {\it vacuum} field equations with
these global properties as presenting a black hole due to a collapse
of gravitational radiations. Clearly, it is an interesting  question
whether such solutions do exist or whether they are excluded by the
field equations.

This question cannot be answered by analysing spherically symmetric
situations. Since the techniques used in
\cite{dafermos:2003} only apply to wave equations in two space-time
dimensions, the answer is not known. If one could perform numerical
calculations which cover the maximal globally hyperbolic solution space-time
one might be able to show the existence of solutions as indicated above. A
natural problem to consider here is the characteristic initial value problem
for the conformal vacuum field equations where data are prescribed on a
cone representing past null infinity (cf. \cite{friedrich:pure rad}). In that
case one would have perfect control on the past but there arise other
difficulties (non-smoothness of the initial hypersurface, how to prescribe
the data on the cone to obtain an asymptotically flat solutions, development
of caustics, the transition of the numerical evolution process through
space-like infinity, etc.) which let the  setting indicated above look more
attractive. 

\vspace{.5cm}

We end this article by pointing out a problem
which did not receive much attention yet but which will become
important as soon as certain technical questions, which are still under
investigation, will be understood.  Depending on the way it
is approached, it poses itself differently,  but it is most severe if
radiation fields are to be calculated numerically. There are
essentially three different approaches to the numerical calculation of
radiation fields: (i) the standard approach based on Cauchy data and
the introduction of an artificial time-like boundary to make the
computational grid finite, (ii) semi-global approaches based either on
characteristic foliations and characteristic initial hypersurfaces extending
to null infinity or on the conformal field equations and hyperboloidal
hypersurfaces, (iii) global approaches, like the one indicated above,
which aspire to calculate entire space-times (possibly including their
asymptotics). In all three cases inner boundaries may be considered to avoid
the approach to singularities, but such boundaries will be ignored here.

In the standard approach particular technical problems are
introduced by the presence of the time-like boundary, which, being in general
not distinguished geometrically, is somewhat unnatural.
Nevertheless, it has been shown by Nagy and Friedrich that the vacuum field
equations admit {\it well posed initial-boundary value problems} (which
includes, of course,  that all constraints be satisfied). There has been
discussed the freedom to prescribe boundary conditions and data and
the pricipal difficulties and specific features of the problem have
been pointed out (\cite{friedrich:nagy}).

It turns out that three real functions can be prescribed on
the time-like boundary. In the setting considered in
\cite{friedrich:nagy} these are the mean extrinsic curvature, which
can be understood as controlling the evolution of the boundary in
the solution space-time (it does not suffice to say `the boundary is
the hypersurface $\{x = 0\}$' for some coordinate $x$), and the
other two are components of the conformal Weyl tensor which may be
interpreted as controlling the two radiative degrees of freedom, though
in general a fully satisfactory physical interpretation of these data
does not exist.  

As pointed out in section \ref{constraints}, a
non-constant mean extrinsic curvature creates problems in the analysis
of the constraints on space-like hypersurfaces. Therefore is is
worth mentioning that the choice of a non-constant mean extrinsic
curvature on the time-like boundary also creates certain difficulties
(cf. \cite{friedrich:nagy}). Though these are quite different from the
ones encountered on space-like hypersurfaces, it seems to indicate
that something important about the mean extrinsic curvature is not
understood yet. We shall not be concerned here with this question,
however, because it does not prevent us from solving the
initial-boundary value problem in all generality.

While the representation of the field equations considered in
\cite{friedrich:nagy} has been used in numerical calculations before
(cf. \cite{frauendiener}), it is not the one used in the majority of
general relativistic numerical codes and there is now a considerable amount
of work being done to derive similar results based on other main evolution
equations (cf. \cite{reula:sarbach}, \cite{sarbach:tiglio} and the
references given there). It can therefore be expected that the basic
numerical problems arising from the initial-boundary value problem will soon
be overcome. Moreover, numerical experiments will also show how the boundary
data must be prescribed to ensure a regular long time evolution of the
boundary.

One will then have to provide a meaningful concept of {\it outgoing
radiation} in an initial-boundary value problem.  A well-defined rigorous
definition is not in sight since in general there does not exist a
distinguished outgoing null vector field transverse to the boundary (the
spherically symmetric case is trivial and in no way representative). We
shall  not be concerned with this question here, pretending that some
answer can be given.  Then there will still remain the question: {\it How
should one dispose of the freedom to prescribe boundary data ?} This problem
has hardly been considered so far and physical intuition is not likely to
give an answer. 

That I am not raising here a purely academic question is illustrated by some
recent calculations.  Allen et al. intend to study in the
article \cite{allen:buckmiller:burko:price}   radiation tails for black
hole evolutions by solving numerically initial boundary value problems
for wave equations on a Schwarzschild background.
The problem is readily reduced to a problem for a wave equation of the
type $\,\,\partial_u\,\partial_v\,\psi = U_l(r)\,\psi$, where $u$ and
$v$ denote the standard retarded and advanced Schwarzschild time
coordinates. The authors prescribe certain initial data
and impose the boundary condition
\begin{equation}
\label{bdrycond}
\partial_v\,\psi = 0 \quad \mbox{on} \quad T = \{r = r_0\},
\end{equation}
for some suitable value $r_0 > 2\,m$ of the standard Schwarzschild
coordinate
$r$. It turns out that radiation tails as predicted by
Price's law (\cite{price}) cannot be identified in the subsequent 
calculations. The authors conclude in their summary: {\it We have shown
that finite-radius boundary conditions prevent the formation of
power-law tails}.

The results of these numerical calculations have been confirmed  
analytically. It has been shown by Dafermos and Rodnianski
(\cite{dafermos:rodnianski:2004}) that 
in the setting of \cite{allen:buckmiller:burko:price} 
any tails vanish on the event horizon faster
than $p(v)^{-1}$ as $v \rightarrow \infty$, where $p$ is any polynomial
of the advanced time coordinate $v$.

One should not think, however,
that these rigorous results confirm the conclusion
above.
Gundlach et al.
successfully verify Price's law numerically by solving characteristic
initial value problems on the same background 
(\cite{gundlach:price:pullin:1994}). This is in accordance with the
analytical results of \cite{dafermos:rodnianski:2003}, where the
characteristic initial value problem is analysed with data prescribed
on a pair of intersecting null hypersurfaces $N_{out}$,
$N_{in}$ which extend to future null infinity and across the horizon
respectively.  

It follows now from the general theory of initial-boundary
value problems that the data given  on $N_{in}$ and the
data induced on a time-like hypersurface $T$ which extends from
$N_{out} \cap N_{in}$ to future time-like infinity determine the
solutions considered in \cite{gundlach:price:pullin:1994} and
\cite{dafermos:rodnianski:2003} uniquely in the future domain of
dependence $D^+$ of $T \cup N_{in}$ (the set of points $p$ for
which any inextendible past directed causal curve through $p$ meets 
$T \cup N_{in}$). There is no reason why the
solution induced on $D^+$ by the numerical results of
\cite{gundlach:price:pullin:1994} should not be reproducible by a
numerical calculation based on the data on $T$ and $N_{in}$. For this
purpose one would have to require the boundary condition
\begin{equation}
\label{secbdrycond}
\partial_v\,\psi = d \quad \mbox{on} \quad T,
\end{equation}
where the function $d$ on $T$ would have to be read off from
the solution given by \cite{gundlach:price:pullin:1994}. 

The calculations in \cite{allen:buckmiller:burko:price} thus do not
indicate that power-law tails cannot be calculated by solving
initial-boundary value problems, they just
confirm that there is a serious problem: {\it In general, one does
not know  the {\it `correct' boundary data} on the right hand side of}
(\ref{secbdrycond}). 

It may be said that the radiation signals one is really interested in
will not be as delicate as the radiation tails. But this does not tell
us how strongly changes in the boundary data will affect wave forms and, in
particular, what will happen in a long time calculation. By a wrong choice of
boundary data, the system which was to be modelled may be affected so
drastically that the approximate radiation field calculated in the end
has little do with the system envisaged originally. To what extent
this can be the case can probably only be explored by numerical
experiments.

The discussion above seems to indicate that the calculation of
wave signals based on characteristic, hyperboloidal, or standard
Cauchy problems will more robust than those based on initial boundary
value problems. But what ever one does, there
will always be an arbitrariness in the choice of data. The work
initiated by Corvino, discussed in section
\ref{constraints}, clearly illustrates the large freedom to deform the
data outside compact sets, or, in other words, how to choose the extension
near space-like infinity of the data which characterize `our system' on a
compact set. 

This leaves one with the task of minimizing the import of
accidental information.
The notion of {\it spurious radiation}  may have an intuitive meaning
but it cannot be well defined.  There are two analytical suggestions,
however, which may supply useful criteria. Dain was able to
associate with a given Cauchy data set
a certain number, related to the {\it global structure} in a
similar sense as the total mass but defined in quite a different way,
which vanishes if and only if the data are stationary
(\cite{dain:2004b}). This number may
thus be considered as a measure for the {\it radiation content} of the
data. It remains to be seen, how changes in the data are reflected in
changes of this number and whether minimizing this quantity in suitable
classes of data leads to applicable results. 

The second suggestion follows from the analysis of asymptotic
smoothness properties. As discussed above, smoothness
requirements at null infinity imply {\it asymptotic conditions}
on the Cauchy data near space-like infinity like asymptotic
staticity or asymptotic stationarity. These may be interpreted as
suppressing `spurious radiation' near space-like infinity.

In any case there will remain some freedom and one needs to assess the effect
of changes of the data near space-like infinity on the wave forms. 
In the end one may
have to look for {\it features of radiation signals which are stable
under the remaining admissible changes of the data}. Identifying 
such features and understanding to what extent they
characterize the type of the source is certainly an important task. It will
require numerical as well as analytical input.


\end{document}